\documentclass[11pt,showpacs,preprintnumbers,amsmath,amssymb]{revtex4}
\usepackage[colorlinks=true,urlcolor=blue,linkcolor=blue]{hyperref}
\usepackage{graphicx}
\usepackage{epsfig}
\usepackage{bm}

\newcommand{\rld}{\rho_{\Lambda D}}
\newcommand{\rl}{\rho_{\Lambda}}
\newcommand{\rlf}{\rho_{\Lambda 5}}

\begin{document}

\title{Restoring Holographic Dark Energy in Brane Cosmology}

\author{E.~N.~Saridakis
\footnote{E-mail: msaridak@phys.uoa.gr}} \affiliation{Department
of Physics, University of Athens, GR-15771 Athens, Greece}

\begin{abstract}
We present a generalized version of holographic dark energy
arguing that it must be considered in the maximally subspace of a
cosmological model. In the context of brane cosmology it leads to
a bulk holographic dark energy which transfers its holographic
nature to the effective 4D dark energy. As an application we use a
single-brane model and we show that in the low energy limit the
behavior of the effective holographic dark energy coincides with
that predicted by conventional 4D calculations. However, a finite
bulk can lead to radically different results.
\end{abstract}

 \maketitle

\section{Introduction}

Holographic dark energy
\cite{Li,hol1,KKEML,Gong,Guberina,Setare,nonflat,Setare2,hdebraneworld}
is an interesting and ingenious idea of explaining the recently
observed Universe acceleration \cite{observ}. Arising from the
cosmological application \cite{holcosm} of the more fundamental
holographic principle \cite{Hooft,witten0}, and despite some
objections on this approach \cite{Linde}, holographic dark energy
reveals the dynamical nature of the vacuum energy by relating it
to cosmological volumes. Its framework is the black hole
thermodynamics \cite{BH,5Dradius} and the connection (known from
AdS/CFT correspondence) of the UV cut-of of a quantum field
theory, which gives rise to the vacuum energy, with the largest
distance of the theory \cite{Cohen}. Such a connection is
necessary for the applicability of quantum field theory in large
distances and results form the argument that the total energy of a
system (which entropy is in general proportional to its volume)
should not exceed the mass of a black hole of the same size (which
entropy is proportional to its area), since in this case the
system would collapse to a black hole violating the second law of
thermodynamics. When this approach is applied to the Universe, the
resulting vacuum energy is identified as holographic dark energy.

Almost all works on the subject remain in the standard 4D
framework. However, brane cosmology, according which our Universe
is a brane embedded in a higher dimensional spacetime
\cite{Rubakov83,RS99b}, apart from being closer to a
higher-dimensional fundamental theory of nature, it has also great
phenomenological successes and a large amount of current research
heads towards this direction \cite{branereview}. It is therefore
desirable to extend holographic dark energy in the braneworld
context. Although there is a recent work dedicated to this goal
\cite{hdebraneworld} it is based on unstable and non-physical
arguments. The main contradiction of the present holographic dark
energy foundation and braneworld models is that although
holographic principle is itself applicable to arbitrary dimensions
\cite{Hooft,Verlinde,iapones} its cosmological application
concerning dark energy should be considered in the maximal
uncompactified space. The reason is that it is the higher
dimensional black hole formation and the higher dimensional
cut-off's which determine the vacuum energy. Therefore, one should
first lay the foundations of bulk holographic dark energy, and
then find the corresponding effective 4D one which appears in
traditional Friedmann equation.

In this work we present this restored holographic dark energy in
brane cosmology. For a specific application we use a general
single-brane model in 4+1 dimensions, although the calculations
can be extended in arbitrary bulk dimensionality. In this
single-brane framework, where the extra dimension is not
restricted, in the low energy limit we recover the results of the
4D holographic dark energy of the literature in flat, open and
closed Universes. We point out that although the obtained
effective 4D behavior coincides with previous works, the
conceptual framework is radically different. This difference
manifests itself in braneworld models with two branes. In the case
of a static bulk with constant interbrane distance, which bounds
the extra dimension \cite{RS99b,twobrane}, the 5D holographic dark
energy and therefore the effective 4D one are constants, losing
their dynamical nature, leaving the case of moving branes as the
only possibility \cite{manos.movingbranes}. The rest of the paper
is organized as follows: In section \ref{HDEBulk} we formulate the
 holographic dark energy in the bulk of general and arbitrary
 spacetimes and in section \ref{HDEBrane} we apply it to a general
 braneworld model in 4+1  dimensions. Finally, in \ref{discussion} we
discuss the physical implications of our analysis and we summarize
the obtained results.

\section{Holographic Dark Energy in the Bulk}\label{HDEBulk}

We consider a general braneworld model where the bulk is
D-dimensional. The restored holographic dark energy states that
the vacuum energy in a volume should not exceed the energy of a
black hole of the same size, both in the maximal subspace, i.e in
the bulk. The mass $M_{BH}$ of a spherical and uncharged
D-dimensional black hole is related to its Schwarzschild radius
$r_s$ through \cite{5Dradius,BHTEV}:
\begin{equation}
M_{BH}=r_s^{D-3}
(\sqrt{\pi}M_D)^{D-3}M_D\frac{D-2}{8\Gamma(\frac{D-1}{2})}.
\label{5drad}
\end{equation}
The D-dimensional Planck mass $M_D$ is related to the
D-dimensional gravitational constant $G_D$ and the usual
4-dimensional Planck mass $M_p$ through:
\begin{eqnarray}
M_{D}=G_D^{-\frac{1}{D-2}}, \nonumber\\
M_p^2=M_D^{D-2}V_{D-4},\label{m5m4}
\end{eqnarray}
where $V_{D-4}$ is the volume of the extra-dimensional space
\cite{5Dradius}.

Now, if $\rld$ is the bulk vacuum energy, then application of the
restored holographic dark energy gives:
\begin{equation}
\rld {\text{Vol}}({\mathcal{S}}^{D-2})\leq r^{D-3}
(\sqrt{\pi}M_D)^{D-3}M_D\,\frac{D-2}{8\Gamma(\frac{D-1}{2})},
\label{resHDE}
\end{equation}
where ${\text{Vol}}({\mathcal{S}}^{D-2})$ is the volume of the
maximal hypersphere in a $D$-dimensional spacetime, given from:
\begin{equation}
{\text{Vol}}({\mathcal{S}}^{D-2})=A_D\,r^{D-1} \label{v2k},
\end{equation}
with
\begin{eqnarray}
A_D=\frac{\pi^{\frac{D-1}{2}}}{\left(\frac{D-1}{2}\right)!}\nonumber,\\
 A_D=\frac{\left(\frac{D-2}{2}\right)!}{\left(D-1\right)!} 2^{D-1}\, \pi^{\frac{D-2}{2}},
\label{ad}
\end{eqnarray}
for $D-1$ being even or odd respectively. Therefore, by saturating
inequality (\ref{resHDE}) introducing $L$ as the largest distance
(IR cut-off) and $c^2$ as a numerical factor, the corresponding
vacuum energy is,  as usual, viewed as holographic dark energy:
\begin{equation}
\rld =c^2 (\sqrt{\pi}M_D)^{D-3}M_D A_D^{-1}
\frac{D-2}{8\Gamma(\frac{D-1}{2})}\,L^{-2} \label{resHD2}.
\end{equation}

Let us make some comments here. Firstly, in general one can obtain
rotational, i.e. non-spherical, or/and charged black holes
\cite{5Dradius,cyllindrical} and furthermore in higher
dimensionality more exotic solutions such as black rings and black
``cigars" are also possible \cite{blackring}. Secondly, the volume
on the left hand side of inequality (\ref{resHDE}) depends on the
specific bulk geometry. However, in order to maintain the
simplicity and universality which lies in the basis of holographic
dark energy, we remain in the aforementioned framework.
Equivalently, the presence of $c^2$ could be regarded as a way of
absorbing all extra numerical factors, thus increasing the
generality of (\ref{resHD2}). Thirdly, let us specify the term
``largest distance'' which was used in the definition of $L$ in
(\ref{resHD2}). Similarly to the usual definition of 4D
holographic dark energy, $L$ could be the Hubble radius, the event
horizon, the square root of the event horizon or the future event
horizon \cite{Li,Guberina,Setare,Hsu,Gong}. For a flat Universe
the last ansatz is the most suitable and furthermore, in this
case, it is the only one that fits holographic statistical
physics, namely the exclusion of those degrees of freedom of a
system that will never be observed by the effective field theory
\cite{Enqvist}. In the majority of braneworld models of the
literature, which are complex and not maximally isotropic in
general, the definition and especially the calculation of the
future event horizon is a hard or impossible task. If the bulk is
finite then the application is retrieved by the use of the volume
of the extra dimensions to define $L$. However, if the extra
dimensions are arbitrary one has to make additional assumptions
for the form of $L$. Lastly, note that a behavior proportional to
$L^{-2}$ was also found in \cite{KKEML} through a different
approach for a special bulk case with a special action term
\cite{Zee} produced by quantum gravity.

A final comment must be made, concerning the sign of bulk
holographic dark energy. In the original Randall-Sundrum model
\cite{RS99b} the bulk cosmological constant should be negative in
order to acquire the correct localization of low-energy gravity on
the brane. Such a negativity is not a fundamental requirement and
is not necessary in more complex, non-static models, especially
when an induced gravity term is imposed explicitly
\cite{inducedgrav}. This is the reason why we include such a term
in the application of the next section, in order to be completely
consistent. However, generally speaking, holographic dark energy
is a simple idea of bounding the vacuum energy from above. It
would be a pity if, despite this effort, one could still have a
negative vacuum energy unbounded from below, because then
holographic dark energy would loose its meaning. If holography is
robust then one should reconsider the case of a negative bulk
cosmological constant (although subspaces, such as branes, could
still have negative tensions). Another possibility is to try to
generalize holographic dark energy to negative values, in order to
impose a negative bound. The subject is under investigation.

\section{Holographic Dark Energy in General 5D Braneworld models}\label{HDEBrane}

In this section we apply the bulk holographic dark energy in
general 5D braneworld models, which constitute the most
investigated case in the field of brane cosmology. We consider an
action of the form \cite{tetradisbulk1,TetradisGB}:
\begin{equation}
 S=\int d^5x\sqrt{-g}\left(M_5^3R-\rlf\right)+\int
 d^4x\sqrt{-\gamma}\,\left({\cal{L}}_{br}^{mat}-V+r_cM^3_5R_4\right).
\label{action}
\end{equation}
In the first integral $M_5$ is the 5D Planck mass, $\rlf$ is the
bulk cosmological constant which is identified as the bulk
holographic dark energy, and $R$ is the curvature scalar of the
5-dimensional bulk spacetime with metric $g_{AB}$. In the second
integral $\gamma$ is the determinant of the induced 4-dimensional
metric $\gamma_{\alpha\beta}$ on the brane, $V$ is the brane
tension and ${\cal{L}}_{br}^{mat}$ is an arbitrary brane matter
content. Lastly, we have allowed for an induced gravity term on
the brane, arising from radiative corrections, with $r_c$ its
characteristic length scale and $R_4$ the 4-dimensional curvature
scalar \cite{inducedgrav,Tetradis.ind.w1}.

In order to acquire the cosmological evolution on the brane we use
the Gaussian normal coordinates with the following metric form
\cite{tetradisbulk1,manos.mirage}:
\begin{equation}
ds^2=-m^2(\tau,y)d\tau^2+a^2(\tau,y)\,d\Omega^2_k+dy^2.
\label{metric}
\end{equation}
The brane is located at $y=0$, we impose a $Z_2$-symmetry around
it, $m(\tau,y=0)=1$ and $d\Omega^2_k$ stands for the metric in a
maximally symmetric 3-dimensional space with $k=-1,0,+1$
parametrizing its spacial curvature. Although we could assume a
general matter-field content \cite{manos.param}, we consider a
brane Universe containing a perfect fluid with equation of state
$p=w\rho$. In this case the low-energy ($\rho\ll V$) brane
cosmological evolution is governed by the following equation
\cite{inducedgrav,TetradisGB}:
\begin{equation}
H^2+\frac{k}{a^2}=\left(72M^6_5+6r_cVM^3_5\right)^{-1}\,V\rho+\frac{V^2}{144M^6_5}+\frac{\rlf}{12M^3_5}.
\label{fried1}
\end{equation}
 In order for equation (\ref{fried1}) to
coincide with the traditional Friedmann equation we have to
impose:
\begin{equation}
V=\frac{72M_5^6}{\frac{3M_p^2}{8\pi}-6r_cM^3_5}. \label{Vm}
\end{equation}
Thus, the evolution of the brane is determined by:
\begin{equation}
H^2+\frac{k}{a^2}=\frac{8\pi\rho}{3M_p^2}+\frac{8\pi\rl}{3M_p^2},
\label{fried2}
\end{equation}
where the (effective in this higher-dimensional model) 4D dark
energy is:
\begin{equation}
\rl\equiv\rho_{\Lambda4}=\frac{M_p^2}{32\pi
M_5^3}\,\rlf+\frac{3M_p^2}{2\pi(\frac{M_p^2}{8\pi M^3_5}-2r_c)^2}.
\label{rl4}
\end{equation}
 In the equations above $\rlf$ is the 5D bulk holographic dark
energy, which according to (\ref{resHD2}) is given by:
\begin{equation}
\rlf=c^2\frac{3}{4\pi}M_5^3L^{-2}\label{rlf}.
\end{equation}
The holographic nature of $\rlf$ is the cause of the holographic
nature of $\rl$. Having in mind that the 5D Planck mass $M_5$ is
related to the standard 4D $M_p$ through $M_5^3=M_p^2/L_5$
(according to (\ref{m5m4})), with $L_5$ the volume (size) of the
extra dimension, we finally acquire the following form for the
effective 4D holographic dark energy:
\begin{equation}
\rl=3c^2\frac{1}{128\pi^2}M_p^2\,L^{-2}+\frac{3M_p^2}{2\pi\left(\frac{L_5}{8\pi}-2r_c\right)^2}.
\label{rl4b}
\end{equation}

The first term in relation (\ref{rl4b}) is a usual holographic
term, with a suitable ansatz for the cosmological length $L$. The
presence of the size $L_5$ of the extra dimension in the second
term accounts for the effects of the restrictions to the
holographic principle by the bulk boundaries. It is obvious that
in general it could radically affect the calculations. In this
work we are interested in investigating the restored holographic
dark energy, without bothering about bulk boundaries, and this is
the justification of the single brane formulation of this section.
Therefore, in the following we assume that $L_5$ is arbitrary
large, much larger than any possible $L$ definition and any other
length scale, thus we omit the second term in relation
(\ref{rl4b}). Its effect will be considered in a separate
publication \cite{manos.movingbranes}.

The final step before the insertion of the 4D effective
holographic dark energy to Friedmann equation (\ref{fried2}) is an
assumption for the cosmological length $L$. As we have already
mentioned, in this work we are interested in presenting the
general bulk holographic dark energy. For the simple application
of this section we will consider a flat Universe, in order to
safely use the future event horizon to define $L$, without
entering into the relevant discussion of the literature concerning
the IR cut-off in non-flat cases
\cite{Li,Guberina,Setare,Hsu,Gong}. However, we stress that bulk
holographic dark energy holds in these cases too, with a suitable
$L$ definition.

The analytical calculation of the future event horizon for the
complete 5D spacetime with metric (\ref{metric}) and dynamics
governed by action (\ref{action}) is impossible. In this
anisotropic model we could alteratively use the 4D future event
horizon, without losing the qualitative behavior of the
observables. Fortunately, the calculations reveal that also the
quantitative results agree with observations and coincide with
those obtained within the traditional holographic dark energy
\cite{Li,hol1,KKEML,Gong,Guberina,Setare,nonflat,Setare2}.

The 4D future event horizon for the FRW brane of our model is as
usual:
\begin{equation}
R_h=a\int_{a}^{\infty}\frac{da'}{Ha'^2}. \label{Rh}
\end{equation}
Inserting this expression to (\ref{rl4b}) we find:
\begin{equation}
\rl=3c^2\frac{1}{128\pi^2}M_p^2\,R_h^{-2}. \label{rl4c}
\end{equation}
Substituting (\ref{rl4c}) to Friedmann equation (\ref{fried2}), in
the flat-Universe case we obtain:
\begin{equation}
\int_{a}^{\infty}\frac{da'}{Ha'^2}=\frac{c}{4\sqrt{\pi}}\left(H^2a^2-\frac{8\pi\rho_0}{3M_p^2a}\right)^{-1/2},
\label{Rh2}
\end{equation}
where we have also introduced the known matter density dependence
on $a$, namely $\rho=\rho_0a^{-3}$, with $\rho_0$ its present
value. The above integral equation determines the brane evolution
and it incorporates the full 5D spacetime effects in low energy
limit, including the bulk holographic dark energy. Seen as 4D
equation in conventional holographic dark energy context it has
been investigated in \cite{Li,nonflat}.

In the case of a flat brane-Universe, one finds the $H$-behavior,
then that of $R_h$ and finally, through (\ref{rl4c}), that of
$\rl$ itself. Inserting the known relation of $\rl$ evolution, i.e
$\rl\sim a^{-3(1+w_\Lambda)}$, and following the steps of
\cite{Li} we find:
\begin{equation}
w_\Lambda=-\frac{1}{3}-\frac{8\sqrt{\pi}}{3c}\sqrt{\Omega_\Lambda^0}+\frac{2\sqrt{\pi}}{3c}\left(1-\Omega_\Lambda^0\right)
\left(\sqrt{\Omega_\Lambda^0}+\frac{8\sqrt{\pi}}{c}\Omega_\Lambda^0\right)z,
\label{wk0}
\end{equation}
where we used $\ln a=-\ln(1+z)\simeq-z$, and $\Omega_\Lambda^0$ is
the present value of $\Omega_\Lambda=8\pi\rl/(3M_p^2H^2)$.
Therefore, according to the value of the constant $c$, one can
obtain a 4D holographic dark energy behaving as phantom
\cite{phantom}, quintessence or quintom \cite{quintom}, i.e
crossing the phantom divide $w=-1$ \cite{w1devide,Tetradis.ind.w1}
during the evolution. Furthermore, one can use observational
results in order to estimate the bounds of the constant
\cite{observHDE1,observHDE}, having in mind that the constants of
every work differ. Finally, one can explore $w_\Lambda$ behavior
in the context of Chaplygin gas or tachyon holographic models
\cite{Setare2}.

\section{Discussion-Conclusions}\label{discussion}

In this work we present holographic dark energy, restored to its
natural foundations, and for a consistency test we apply it to a
general braneworld model well studied in the literature. Our main
motivation is to match the successes of brane cosmology in both
theoretical and phenomenological-observational level, with the
successful, simple, and inspired by first principles, notion of
holographic dark energy in conventional 4D cosmology. Our basic
argument for this generalization is that in a higher dimensional
spacetime, it is the bulk space which is the natural framework for
the cosmological application, concerning dark energy, of
holographic principle, and not the lower-dimensional
brane-Universe. This is obvious since it is the
maximally-dimensional subspace that determines the properties of
quantum-field or gravitational theory, and this holds even if we
consider brane cosmology as an intermediate limit of an even
higher dimensional fundamental theory of nature. To be more
specific we recall here that the underlying idea of holographic
dark energy is that one cannot have more energy in a volume than
the mass of a black hole of the same size. In braneworld models,
where the spacetime dimension is more than 4, black holes will in
general be D-dimensional \cite{BH,5Dradius}, no matter what their
4D effective (mirage) effects could be. Therefore, although
holography itself can be applied in arbitrary subspaces, its
specific cosmological application which eventually gives rise to
holographic dark energy holds only for the main space, i.e the
bulk. Subsequently, this bulk holographic dark energy will bring
forth an effective 4D dark energy with ``inherited" holographic
nature, and this one will be present in the Friedmann equation of
the brane. Completing these self-consistent cogitations, the brane
Friedmann equation should arise from the full dynamics, too.

The generalized holographic dark energy can lead to either
radically different or exactly identical 4D behavior, comparing to
that obtained in conventional 4D literature
\cite{Li,hol1,KKEML,Gong,Guberina,Setare,nonflat,Setare2}. In
section \ref{HDEBulk} we applied this bulk holographic dark energy
in a general braneworld model, with an induced gravity term and a
perfect fluid on the brane. In this case, the low energy evolution
of the single brane leads to effective 4D holographic dark energy
behaving either as a phantom, quintessence or quintom, identically
to conventional 4D calculations. However, as we have mentioned, it
is obvious that the interpretation and explanation of this
behavior is fundamentally different. The reason for the coinciding
results in this specific example is that the extra dimension is
arbitrary large, imposing no restrictions on the application of
holographic dark energy. The only necessary assumption of our
calculation is the use of the 4D future event horizon as the
cosmological length of holographic behavior. This can be justified
since in this anisotropic model the 5D event horizon cannot be
found analytically, and furthermore, under the requirement of the
recovery of conventional evolution on the brane, in agreement with
observations, the 5D future event horizon cannot be significantly
different than its 4D counterpart. Note however, that in other
braneworld models, where the whole space is FRW, dS or AdS
\cite{KKEML,iapones}, the D-dimensional future event horizon can
be easily calculated, and this would lead to an exact application
of the bulk holographic dark energy.

From this discussion it becomes evident that a finite bulk would
radically affect the 4D behavior of dark energy. For example, a
two brane model with constant interbrane distance would lead to a
constant dark energy, and the physical interpretation is that it
is impossible to have an arbitrary large bulk black hole in this
case. However, one could built a two-brane model where the branes
moving apart. The subject is under investigation
\cite{manos.movingbranes}.

In this work we present a restored version of holographic dark
energy in brane cosmology, and we argue that it has to be
considered in the bulk and not in the brane. As an example we use
a single-brane model and we show that, in the low energy limit,
the 4D effective holographic dark energy behavior coincides with
that predicted by conventional 4D calculation. However, its
behavior in more complicated braneworld models can be
significantly different.
\\

\paragraph*{{\bf{Acknowledgements:}}}
The author is grateful to  G.~Kofinas, K.~Tamvakis, N. Tetradis,
F. Belgiorno, B. Brown, S. Cacciatori, M. Cadoni, R.~Casadio,
G.~Felder, A.~Frolov, B. Harms, N.~Mohammedi, M.~Setare and
Y.~Shtanov for useful discussions.

\end{document}